\documentclass[a4paper,12pt]{article}

\usepackage[font={small}]{caption}
\usepackage{amsmath}
\usepackage{color,graphicx}
\usepackage[top=2.5cm,right=2cm,bottom=2.5cm,left=2cm]{geometry}
\usepackage[colorlinks,linkcolor=blue,citecolor=blue,pagebackref=false]{hyperref}
\usepackage{cite}
\usepackage[parfill]{parskip}
\usepackage{multirow}
\usepackage{xurl}
\usepackage{subcaption}

\begin{document}

\begin{center}

\textbf{Phase- and amplitude-dependent control of synchronization in excitatory-inhibitory networks via pulsed stimulation}

\vspace{0.5 cm}

Ehsan Ahmadi\textsuperscript{1},
Mojtaba Madadi Asl\textsuperscript{2,3}, and %\let\thefootnote\relax\footnotetext{* Email address: m.madadi@ipm.ir.}
Alireza Valizadeh\textsuperscript{1,4,*} \let\thefootnote\relax\footnotetext{* Corresponding author. Email address: valizadeh@gmail.com.}
\\
\bigskip
\footnotesize{
\textsuperscript{1}Department of Physics, Institute for Advanced Studies in Basic Sciences (IASBS), Zanjan, Iran\\
\textsuperscript{2}School of Biological Sciences, Institute for Research in Fundamental Sciences (IPM), Tehran, Iran\\
\textsuperscript{3}Pasargad Institute for Advanced Innovative Solutions (PIAIS), Tehran, Iran\\
\textsuperscript{4}The Zapata-Briceño Institute of Neuroscience, Madrid, Spain\\
}

\end{center}

%%%%%%%%%%%%%%%%%%%%%%%%%%%%%%%%%%%%%%%%%%%%%%%%%%%%%
%%%%%%%%%%%%%%%%%%%%%%%%%%%%%%%%%%%%%%%%%%%%%%%%%%%%%

\vspace{0.4cm}

\begin{center}

%(Last modified on: \today)
%\vspace{0.4cm}

{\small

\begin{minipage}{14cm}

\begin{center}
\textbf{Abstract}
\end{center}

\vspace*{0.2cm}

Oscillatory neuronal networks exhibit complex collective responses to external perturbations that depend on both the intrinsic network dynamics and the timing of stimulation. Although phase response curves (PRCs) have become a standard tool for characterizing these responses, phase resetting alone provides an incomplete description of how transient perturbations reshape collective activity. Here, we investigate the dynamics of a balanced excitatory-inhibitory network of exponential integrate-and-fire (EIF) neurons subjected to phase-targeted current pulses. By jointly analyzing the network phase response curve (nPRC), network amplitude response curve (nARC), and changes in the population synchrony, we establish a framework for characterizing collective network responses in terms of phase, amplitude, and synchronization. We show that identical stimulation pulses can either enhance, suppress, or leave network synchronization unchanged depending solely on their phase within the oscillation cycle, revealing distinct synchronizing and desynchronizing windows. The nARC further identifies robust phase intervals that maximize suppression of oscillatory activity and provide optimal targets for repeated stimulation. Successive perturbations progressively desynchronize the network activity while continuously reshaping the phase, amplitude, and synchrony response landscapes, driving the network toward a modified dynamical state without altering the optimal stimulation phase. These cumulative effects remain robust across stimulation intensities, inhibitory synaptic time constants, and independent network realizations. Our results demonstrate that phase, amplitude, and synchronization represent complementary dynamical dimensions of oscillatory neuronal networks and provide general principles for the state-dependent control of collective dynamics through phase-targeted perturbations.\\


\end{minipage}
}
\end{center}

%%%%%%%%%%%%%%%%%%%%%%%%%%%%%%%%%%%%%%%%%%%%%%%%%%%%%
%%%%%%%%%%%%%%%%%%%%%%%%%%%%%%%%%%%%%%%%%%%%%%%%%%%%%

%\newpage
\vspace{0.5cm}

\section{Introduction}

Synchronized oscillations are a fundamental phenomenon in neuronal networks, where coordinated activity of neurons underlies both normal brain functioning and pathological states. In cortical networks, excitatory-inhibitory interactions are known to generate oscillations in beta (15-30 Hz) and gamma (30-100 Hz) bands which are associated with a wide range of cognitive processes such as perception, attention, memory, and motor control~\cite{bacsar2001gamma,ward2003synchronous,buzsaki2004neuronal,fries2005mechanism,wang2010neurophysiological}. However, abnormal oscillatory activity is implicated in several brain disorders such as Parkinson's disease, schizophrenia, and epilepsy~\cite{dominguez2005enhanced,uhlhaas2006neural,hammond2007pathological,uhlhaas2010abnormal}. Therefore, understanding and controlling synchronization in neuronal systems is of both theoretical and practical importance.
 
External stimulation has emerged as a powerful tool for modulating neuronal dynamics~\cite{madadiasl2025entrainment}. In particular, pulsed stimulation techniques for the modulation of synchronized activity are widely used in both computational models~\cite{tass2001desynchronizing,tass2003model,rubin2004high,holt2016phasic,madadi2023decoupling} and clinical applications~\cite{paus2001synchronization,meissner2005subthalamic,kuhn2008high,thut2011rhythmic}, such as deep brain stimulation (DBS) or transcranial magnetic stimulation (TMS). However, the effectiveness of brain stimulation interventions, in general, depends sensitively on the timing, duration, intensity, and frequency of stimulation~\cite{zhao2024intensity}. For example, several studies have shown that stimulation precisely timed to the phase of pathologically synchronized oscillations can restore healthy network states~\cite{cagnan2017stimulating,holt2019phase,duchet2020phase,west2022stimulating}. A key challenge is to determine how to optimally design stimulation protocols to achieve desired dynamical outcomes.

One of the most established tools for analyzing the response of oscillatory systems to perturbations is the phase response curve (PRC), which quantifies how a transient perturbation shifts the phase of an oscillator, providing crucial insights into the controllability of synchronized activity states~\cite{hansel1995synchrony,ermentrout1996type}. While PRCs have been extensively studied at the single-neuron level, extending this framework to population dynamics, referred to as network PRC (nPRC)~\cite{kawamura2008collective,levnajic2010phase,dumont2017macroscopic}, remains nontrivial due to nonlinear interactions and heterogeneity within networks. In addition, the network amplitude response curve (nARC)~\cite{aronson1990amplitude}, which captures changes in oscillation magnitude due to brief perturbations, provides complementary information that is often overlooked. In fact, both phase and amplitude responses are crucial for understanding how external stimuli influence ongoing oscillations in neuronal networks~\cite{smeal2010phase}.

In networks of interacting neurons, collective oscillations arise from the dynamic interplay between excitation and inhibition, reflecting the coordinated activity of large neuronal populations~\cite{van1994inhibition,whittington2000inhibition,tiesinga2009cortical}. These interactions can either enhance, stabilize, or disrupt neuronal synchrony depending on factors such as network topology, synaptic strengths, conduction delays, and intrinsic neuronal properties~\cite{brunel2000dynamics,ledoux2011dynamics,mejias2014differential,brito2021neuronal}. Consequently, the response of neuronal networks to external perturbations cannot be fully characterized by phase resetting alone. External stimuli may simultaneously alter the timing, amplitude, and coherence of ongoing oscillations, thereby reshaping network dynamics in multiple ways. Understanding these effects is particularly important for stimulation-based interventions, where identical stimuli may produce markedly different outcomes depending on the instantaneous state of the network. A comprehensive description of network responses therefore requires jointly considering phase resetting, amplitude modulation, and their consequences for collective synchronization. This multidimensional perspective provides a more complete characterization of how external perturbations reshape oscillatory network dynamics than phase-based analyses alone.

In this study, we investigate how transient current pulses reshape the collective dynamics of a balanced excitatory-inhibitory network of exponential integrate-and-fire (EIF) neurons. We first characterize the spontaneous oscillatory state of the network and then quantify its response to phase-targeted perturbations using the nPRC and nARC. To complement these measures, we evaluate changes in the population Fano factor (pFF), allowing us to directly assess how external stimulation modifies collective synchronization. By jointly analyzing these three quantities, we describe network responses in terms of phase resetting, amplitude modulation, and synchronization. Specifically, characteristics of network responses can be employed to manipulate oscillatory activity by external perturbations.

Accordingly, we identify distinct phase windows in which identical stimulation pulses either enhance, suppress, or minimally affect network synchrony. We further demonstrate that repeated stimulation delivered within the desynchronizing window progressively weakens oscillatory activity and drives the network toward a less synchronized dynamical state, while preserving the overall phase dependence of the response. Importantly, these cumulative effects remain robust across variations in stimulation intensity, inhibitory synaptic time constants, and independent network realizations. Together, these findings reveal that phase, amplitude, and synchronization constitute complementary dimensions of collective network dynamics and provide general principles for designing phase-specific stimulation strategies to control pathological oscillations in oscillatory networks.

%%%%%%%%%%%%%%%%%%%%%%%%%%%%%%%%%%%%%%%%%%%%%%%%%%%%%
%%%%%%%%%%%%%%%%%%%%%%%%%%%%%%%%%%%%%%%%%%%%%%%%%%%%%

\section{Methods}
\subsection{Network and neuron model}

\begin{figure}[t!]
\centering
\includegraphics[scale = 0.9]{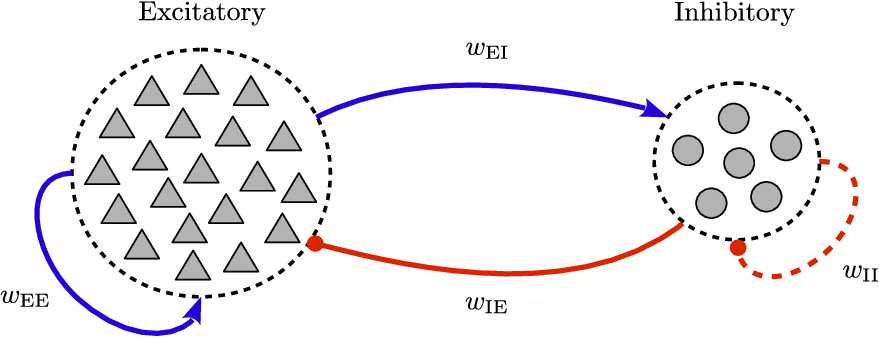}
\caption{{\bf Representation of the network model.} The model consists of a network of $N_{\mathrm{E}} = 800$ excitatory (triangles) and $N_{\mathrm{I}} = 200$ inhibitory (circles) neurons randomly connected via conductance-based synapses. The excitatory and inhibitory populations in the model interact via synaptic weights $w_{\rm EI}$ (E $\rightarrow$ I) and $w_{\rm IE}$ (I $\rightarrow$ E). Within-population interactions are mediated by synaptic weights $w_{\rm EE}$ (E $\rightarrow$ E) and $w_{\rm II}$ (I $\rightarrow$ I).}
\label{fig1}
\end{figure}

As schematically shown in Fig.~\ref{fig1}, we considered a network of $N = 1000$ EIF neurons, consisting of $N_{\mathrm{E}} = 800$ excitatory and $N_{\mathrm{I}} = 200$ inhibitory units in 4:1 proportion. Synaptic interactions were modeled with sparse random connectivity and fixed connection probability of $p = 0.1$. The excitatory and inhibitory populations were coupled via synaptic weights $w_{\rm EI}$ (E $\rightarrow$ I) and $w_{\rm IE}$ (I $\rightarrow$ E). Within-population interactions were mediated by synaptic weights $w_{\rm EE}$ (E $\rightarrow$ E) in the excitatory population and $w_{\rm II}$ (I $\rightarrow$ I) in the inhibitory population. The membrane potential ($V_{\rm m}$) dynamics govern~\cite{fourcaud2003spike}:
\begin{equation}\label{eq:1}
C_{\rm m} \frac{dV_{\rm m}}{dt} = - g_{\rm L} (V_{\rm m} - E_{\rm L}) + g_{\rm L} \Delta_T \exp \left( \dfrac{V_{\rm m} - V_{\rm th}}{\Delta_T} \right)  + I_{\rm syn}(t) + I_{\rm ext},
\end{equation}
where $C_{\rm m}$ is the membrane capacitance, $g_{\rm L}$ is the leak conductance, $E_{\rm L}$ is the leak reversal potential, and $\Delta_T$ is the slope factor for spike initiation. The model includes a threshold for spike initiation ($V_{\rm th}$), and a fixed refractory period ($t_{\rm ref}$). A spike is emitted at time step $t^* = t_{k + 1}$ if $V_{\rm m}(t_k) < V_{\rm th}$ and $V_{\rm m}(t_{k + 1}) > V_{\rm th}$. After that, the membrane potential resets to $V_{\rm{reset}}$ for $t^* < t <  t^* + t_{\rm ref}$, i.e., the membrane potential is clamped to $V_{\rm{reset}}$ during the refractory period. The numerical values of model parameters are listed in Table~\ref{table1}.

%%%%%%%%%%%%%%%%%%%%%%%%%%%%%%%%%%%%%%%%%%%%%%%%%%%%%

\subsection{Synapse model}

In the model, neurons were coupled in a random topology via conductance-based synapses, with synaptic interactions were realized by the synaptic current $I_{\rm syn}(t)$ as introduced in Eq.~(\ref{eq:1}). The total synaptic current received by a neuron comprised excitatory and inhibitory components, i.e., $I_{\rm syn}(t) = I^{\rm ex}_{\rm syn}(t) + I^{\rm in}_{\rm syn}(t)$, where each component is given by:
\begin{equation}\label{eq:2}
I^{\rm X}_{\rm syn}(t) = (E^{\rm X}_{\mathrm{syn}} - V_{\rm m}(t)) \sum_{i} \sum_{f} g_{i \rm X}(t - t^{(f)}_i - d_{\rm syn}),
\end{equation}
where $i$ represents excitatory (X = E) or inhibitory (X = I) presynaptic neurons characterized by the membrane potential $V_{\rm m}(t)$, $E^{\rm X}_{\mathrm{syn}}$ is the corresponding synaptic reversal potential, $g(t)$ is the synaptic conductance, $t^{(f)}$ is the spike times of the presynaptic neuron, and $d_{\rm syn}$ is the synaptic delay. Unlike models with instantaneous rise times, each incoming spike produces a postsynaptic conductance change described by an alpha function. The resulting synaptic conductance profile for a single presynaptic spike is given by: 
\begin{equation}\label{eq:3}
g_{i \rm X}(t) = w_i \left(\dfrac{t}{\tau^{\rm X}_{\rm{syn}}}\right) \exp \left(1 - \dfrac{t}{\tau^{\rm X}_{\rm{syn}}} \right)  \Theta(t),
\end{equation}
where $w$ is the synaptic weight, $\tau^{\rm X}_{\rm{syn}}$ is the synaptic time constant, and $\Theta(t)$ denotes the Heaviside step function, defined as $\Theta(t) = 1$ for $t \geq 0$, and $\Theta(t) = 0$ otherwise. The synaptic conductances are normalized to unit maximum~\cite{meffin2004analytical}, i.e., $g (t = 0) = w$.

%%%%%%%%%%%%%%%%%%%%%%%%%%%%%%%%%%%%%%%%%%%%%%%%%%%%%

%%%%%%%%%%%%%%%%%%%%%%%%%% Fig1
%%%%%%%%%%%%%%%%%%%%%%%%%%

%%%%%%%%%%%%%%%%%%%%%%%%%%%%%%%%%%%%%%%%%%%%%%%%%%%%%

\subsection{External background input}

Instead of employing discrete Poisson spike trains to model background activity, the baseline dynamical state of the network in our simulations is established through continuous somatic current injections, defined as $I_{\rm ext}$ in Eq.~(\ref{eq:1}). This approach mimics the net effect of massive, uncorrelated synaptic bombardment from external cortical sources. Each neuron receives a constant external direct current, defined as $700$ pA, which continuously depolarizes the membrane potential towards the exponential spike-initiation threshold. To incorporate the inherent stochasticity of \textit{in-vivo}-like conditions and intrinsic membrane fluctuations of cortical neurons~\cite{shadlen1998variable}, a Gaussian white noise current with a standard deviation of $\sigma =7.0$ pA is superimposed onto the deterministic baseline drive. This configuration provides a robust and computationally highly controllable background drive, allowing for precise isolation of the network's intrinsic oscillatory dynamics prior to external pulsing.

%%%%%%%%%%%%%%%%%%%%%%%%%%%
\begin{table}[t!]
\centering
\caption{{\bf Parameters of the network and neuron model used in our simulations.}}
\vspace*{-0.2cm}
\begin{tabular}{|l|c|c|c|}
\hline
{\bf Parameter} & {\bf Symbol} & {\bf Value} & {\bf Unit}\\ \hline
Membrane capacitance & $C_{\rm m}$ & 250 & $ \mathrm{pF}$\\ \hline
Spiking threshold & $V_{\mathrm{th}}$ & -50 & $\mathrm{mV}$\\ \hline
Resting membrane potential & $V_{\mathrm{reset}}$ & -68 & $\mathrm{mV}$\\ \hline
Leak reversal potential & $E_{\rm L}$ & -70 & $\mathrm{mV}$\\ \hline
Leak conductance & $g_{\rm L}$ & 30 & $\mathrm{nS}$\\ \hline
Duration of refractory period & $t_{\rm ref}$ & 1.7 & $\mathrm{ms}$\\ \hline
Slope factor & $\Delta_T$ & 2.0 & $\mathrm{mV}$\\ \hline
Excitatory reversal potential & $E^{\mathrm{ex}}_{\rm syn}$ & 0 & $\mathrm{mV}$\\ \hline
Inhibitory reversal potential & $E^{\mathrm{in}}_{\rm syn}$ & -80 & $\mathrm{mV}$\\ \hline
Decay time constant of excitatory synapses & $\tau^{\rm ex}_{\rm{syn}}$ & 2.0 & $\mathrm{ms}$\\ \hline
Decay time constant of inhibitory synapses & $\tau^{\rm in}_{\rm{syn}}$ & 6.5 & $\mathrm{ms}$\\ \hline
Synaptic delay & $d_{\rm syn}$ & 2.0 & $\mathrm{ms}$\\ \hline
${\rm E} \rightarrow {\rm E}$ synaptic weight & $w_{\rm{EE}}$ & 0.1 & $\mathrm{nS}$\\ \hline
${\rm E} \rightarrow {\rm I}$ synaptic weight & $w_{\rm{EI}}$ & 0.1 & $\mathrm{nS}$\\ \hline
${\rm I} \rightarrow {\rm E}$ synaptic weight & $w_{\rm{IE}}$ & -0.49 & $\mathrm{nS}$\\ \hline
${\rm I} \rightarrow {\rm I}$ synaptic weight & $w_{\rm{II}}$ & -0.49 & $\mathrm{nS}$\\ \hline
Number of excitatory neurons & $N_{\mathrm{E}}$ & 800 & -- \\ \hline
Number of inhibitory neurons & $N_{\mathrm{I}}$ & 200 & -- \\ \hline
Total number of neurons & $N$ & 1000 & -- \\ \hline
Connection probability & $p$ & 0.1 & -- \\ \hline
\end{tabular}
\label{table1}
\end{table}
%%%%%%%%%%%%%%%%%%%%%%%%%%%

%%%%%%%%%%%%%%%%%%%%%%%%%%%%%%%%%%%%%%%%%%%%%%%%%%%%%

\subsection{Phase and amplitude responses to external stimuli}

The PRC of a periodically spiking neuron characterizes how brief perturbations influence the timing of subsequent spikes~\cite{hansel1995synchrony,ermentrout1996type}. Typically, this is measured by applying a small current pulse and quantifying the resulting shift in spike timing or phase of the oscillating neuron. To investigate the response of the excitatory-inhibitory network to external perturbations in our model, we employed the nPRC framework introduced previously~\cite{kawamura2008collective,levnajic2010phase,dumont2017macroscopic}, which extends the single-neuron PRC to the population level. The nPRC captures the collective phase shift of network oscillations in response to external inputs, thereby providing a macroscopic measure of the network's sensitivity to perturbations.

To quantify the nPRC, we first adjusted the model parameters so that the unperturbed network exhibited synchronized activity. The collective oscillation cycle ($T_0$) was estimated by measuring the interval between two successive peaks of the network oscillation. The network was then perturbed by injecting a brief, rectangular positive current pulse with a duration of 2 ms (and different intensities) at different phases ($\theta$) of the oscillation cycle. The resulting phase shifts of the network oscillation were measured by:
\begin{equation}\label{eq:4}
{\rm nPRC}(\theta) = 1 - \dfrac{T^{\prime}}{T_0},
\end{equation}
where $\theta = \frac{2 \pi}{T_0} (t  - t_n)$ is the phase of network oscillation, $T^{\prime} = t_{n + 1} - t_{n}$ is the oscillation cycle of the perturbed network, $t_{n}$ is the time of the last synchronous population peak before perturbation and $t_{n + 1}$ is the time of the first synchronous population peak after perturbation that could be advanced or delayed.

%To map the state-dependent responses, the stimulation protocol utilized varying pulse intensities, specifically utilizing $I_1 = 70$ pA (strong) and $I_2 = 10$ pA (weak). These perturbations were selectively applied to targeted subsets of the neuronal population (e.g., 10 neurons). The stimuli were systematically delivered across uniformly distributed phases of the ongoing population oscillation cycle (e.g., 10 distinct phase points).

To quantify the network's amplitude response, we computed the nARC~\cite{aronson1990amplitude}, providing a measure of how temporally localized perturbations modulate the magnitude of network oscillations at the population level. Specifically, for each perturbation, we measured the change in amplitude caused by the perturbation:
\begin{equation}\label{eq:5}
{\rm nARC}(\theta) = \dfrac{A^{\prime}}{A_0} - 1,
\end{equation}
where $A_0 = A(t_n)$ is the unperturbed peak amplitude and $A^{\prime} = A(t_{n + 1})$ is the amplitude of the first synchronous population peak after perturbation. By computationally tracking the asymptotic trajectory of the macroscopic variables post-stimulation, we extracted the phase deviations and amplitude variations, thereby constructing the nPRC and nARC essential for predicting the network's response to multi-pulse protocols. 

%%%%%%%%%%%%%%%%%%%%%%%%%%%%%%%%%%%%%%%%%%%%%%%%%%%%%

\subsection{Analysis of the network dynamics}
\subsubsection{Population activity}

To evaluate the macroscopic synchronized dynamics of the network model, the population activity was computed for both excitatory and inhibitory subpopulations, excluding the first 1500 ms of transients:
\begin{equation}\label{eq:6}
A_{\rm X}(t) = \dfrac{1}{N_{\rm X}} \sum^N_{i=1} \sum_f \delta(t - t_i^{(f)}),
\end{equation}
where $N_{\rm X}$ is the total number of neurons in the excitatory ($N_{\mathrm{E}} = 800$) or inhibitory ($N_{\mathrm{I}} = 200$) subpopulation, and $t^{(f)}$ represents the firing time of individual neurons. $A(t)$ is reported as the percentage of active neurons.

%%%%%%%%%%%%%%%%%%%%%%%%%%%%%%%%%%%%%%%%%%%%%%%%%%%%%

\subsubsection{Synchrony index}

The degree of coherent firing and synchronization across the network was quantified using the pFF, excluding the first 1500 ms of transients~\cite{kumar2008conditions}:
\begin{equation}\label{eq:7}
{\mathrm{pFF}} = \dfrac{\sigma^2 [A(t)]}{\mu [A(t)]},
\end{equation}
where $A(t)$ represents the population activity defined in Eq.~(\ref{eq:6}), and $\sigma^2$ and $\mu$ are the variance and mean of the population activity, respectively. A completely asynchronous (Poisson-like) network state yields a pFF close to unity, whereas highly synchronized oscillatory states are characterized by significantly larger pFF values due to extensive amplitude fluctuations in population activity~\cite{kumar2008conditions,madadi2023decoupling}. Smaller values of the pFF correspond to desynchronized states, whereas greater values of the pFF imply synchrony in the network.

%%%%%%%%%%%%%%%%%%%%%%%%%%%%%%%%%%%%%%%%%%%%%%%%%%%%%

\subsubsection{Spike count irregularity}

The coefficient of variation (CV) of the inter-spike intervals (ISIs) was calculated to quantify the temporal irregularity of the spiking activity of neuron $i$, as follows:
\begin{equation}\label{eq:8}
{\mathrm{CV}_i} = \dfrac{\sigma_i}{\mu_i},
\end{equation}
where $\sigma_i$ is the standard deviation and $\mu_i$ is the mean of the ISIs distribution for the $i$-th neuron. CV values approaching 1 imply a highly irregular, Poisson-like firing pattern, whereas values significantly lower than 1 are indicative of regular spiking behavior. The population-averaged CV was utilized to assess the microscopic stochasticity of the network state.

%%%%%%%%%%%%%%%%%%%%%%%%%%%%%%%%%%%%%%%%%%%%%%%%%%%%%

\subsubsection{Network frequency and power spectrum}

The macroscopic oscillatory behavior of the network was characterized by extracting the dominant frequency from the power spectral density (PSD) of the population activity. The PSD was estimated using Welch's method~\cite{welch1967use}, employing overlapping segments (50\% overlap) and using a sampling frequency of 1000 Hz. The network oscillation frequency was strictly defined as the frequency corresponding to the global peak of the computed PSD to identify dominant oscillatory states under varying computational regimes.

%%%%%%%%%%%%%%%%%%%%%%%%%%%%%%%%%%%%%%%%%%%%%%%%%%%%%

\subsection{Model integration}

Computer simulations were performed using the NEural Simulation Tool (NEST)~\cite{gewaltig2007nest}. The simulation code and the associated data used to generate the figures are publicly accessible from \url{https://github.com/ehsanahmadi0013/Desynchronizing-phase-stimulation}. The subthreshold dynamics were integrated using the \texttt{aeif\_cond\_alpha} model implemented in NEST originally designed to simulate adaptive exponential integrate-and-fire (AdEx) neurons~\cite{brette2005adaptive}. To ensure the model acts as a pure EIF model with no adaptation mechanisms, both subthreshold ($a$) and spike-triggered ($b$) adaptation parameters were explicitly set to zero. All other unspecified intrinsic neuronal parameters were kept at their NEST default values. The network architecture consists of a randomly connected, sparse topology of these conductance-based EIF neurons. The underlying differential equations were integrated using a temporal resolution $dt = 0.01$ ms. Each numerical experiment was simulated for a total duration of $T = 3000$ ms, corresponding to $3.0 \times 10^5$ integration steps. Data analysis and parameter explorations were executed using customized Python scripts interfacing with PyNEST~\cite{eppler2009pynest}.

%%%%%%%%%%%%%%%%%%%%%%%%%%%%%%%%%%%%%%%%%%%%%%%%%%%%%
%%%%%%%%%%%%%%%%%%%%%%%%%%%%%%%%%%%%%%%%%%%%%%%%%%%%%

\section{Results}
\subsection{Baseline network dynamics}

\begin{figure}[t!]
\centering
\includegraphics[scale = 0.3]{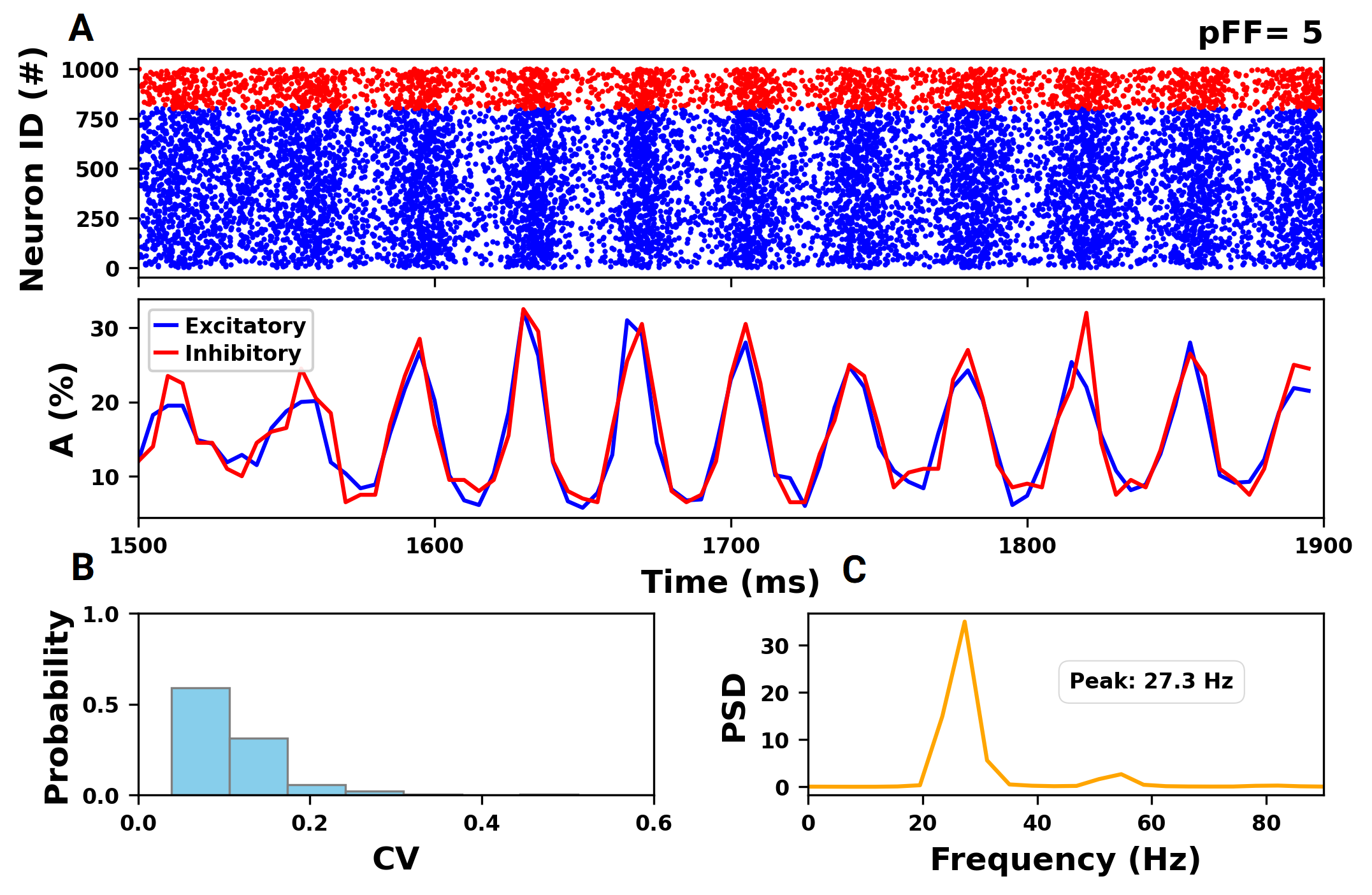}
\caption{{\bf Dynamics of excitatory-inhibitory network.} (\textbf{A}) Raster plot (top) and population activity (bottom) of the network of excitatory (blue) and inhibitory (red) neurons. Model parameters are listed in Table~\ref{table1}. The level of synchrony (assessed by calculating pFF) is indicated above the panel (pFF = 5.0). (\textbf{B}) Spike count irregularity of individual neurons (assessed by calculating CV). (\textbf{C}) PSD of the population activity, showing an oscillatory peak at approximately 27.3 Hz.}
\label{fig2}
\end{figure}

The dynamics of the baseline (unperturbed) excitatory-inhibitory network, as illustrated in Fig.~\ref{fig2}, reveal a balanced state of activity where excitatory (blue) and inhibitory (red) neurons interact to produce structured population patterns. The raster plot (Fig.~\ref{fig2}A, top) shows that spikes from excitatory and inhibitory neurons are distributed across the network with some degree of temporal clustering, as reflected in population activity (Fig.~\ref{fig2}A, bottom). The calculated pFF as a measure of synchronized activity (pFF = 5.0) indicates moderate synchrony at the population level, suggesting that while individual neurons fire irregularly, the network exhibits coordinated fluctuations in overall activity.

Individual neuron spiking irregularity, assessed through the CV in Fig.~\ref{fig2}B, shows that most neurons display irregular firing, with CV values primarily below 0.2. This indicates that neurons do not fire at perfectly regular intervals but instead exhibit a stochastic pattern consistent with in vivo cortical firing~\cite{shadlen1998variable}. Such variability at the single-cell level supports robust network computation by preventing overly predictable dynamics and allowing the network to respond flexibly to inputs. The spectral analysis of the network's population activity (Fig.~\ref{fig2}C) reveals a pronounced oscillatory peak at approximately 27.3 Hz, highlighting a preferred frequency of collective network oscillations. This peak is consistent with low-gamma/fast-beta range oscillatory activity often observed in cortical circuits, reflecting the interplay between excitation and inhibition. These baseline network dynamics demonstrate a balanced and irregularly firing neuronal population that supports coherent population-level oscillations, setting the stage for further investigation of network responses under varied input conditions.

%%%%%%%%%%%%%%%%%%%%%%%%%% Fig2
%%%%%%%%%%%%%%%%%%%%%%%%%%

%%%%%%%%%%%%%%%%%%%%%%%%%%%%%%%%%%%%%%%%%%%%%%%%%%%%%

\subsection{Phase-dependent response to single pulses}

\begin{figure}[t!]
\centering
\includegraphics[scale = 0.4]{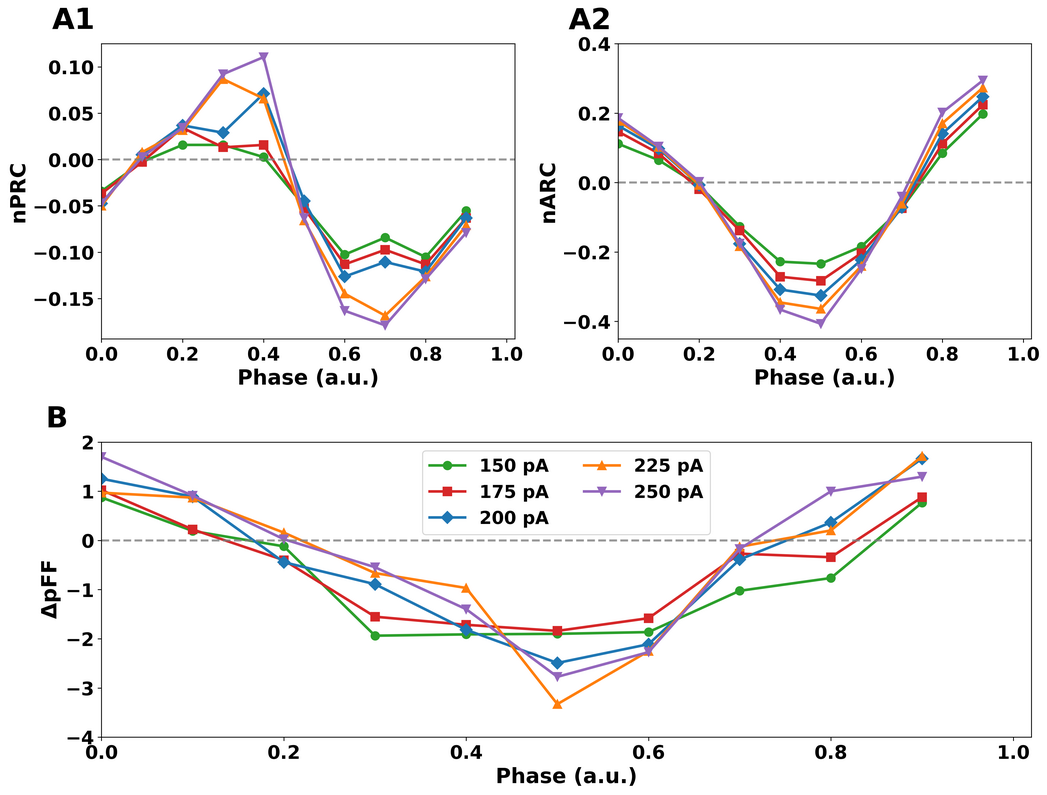}
\caption{{\bf Population phase and amplitude responses to single pulses.} (\textbf{A1, A2}) The nPRC (A1) and nARC (A2) of the baseline excitatory-inhibitory network obtained by perturbing the whole population at different phases of the oscillation cycle (see Methods). The gray dashed lines indicate zero response. (\textbf{B}) Change in the pFF level in the network receiving single-pulse stimuli (the perturbed network) compared to
its corresponding baseline (unperturbed) network, defined as $\rm \Delta pFF = pFF_{Perturbed} - pFF_{Baseline}$. In the figure, curves with different colors and markers belong to different pulse intensities. The gray dashed line denotes zero change.}
\label{fig3}
\end{figure}

We first examine the response of the network to single pulses applied at different phases of the oscillation cycle (see Methods). The phase-dependent response analysis of the baseline network (Fig.~\ref{fig3}) shows that perturbations induce markedly different effects depending on their timing within the cycle. The nPRC (Fig.~\ref{fig3}A1) exhibits a consistent biphasic structure across input intensities: early-phase stimulation tends to delay the next cycle (negative nPRC values), whereas mid-to-late phase stimulation generally advances the oscillation (positive nPRC values). Importantly, the magnitude of these phase shifts increases with pulse intensity, with stronger inputs (225 pA and 250 pA) producing more pronounced advances and delays, indicating a graded control of phase resetting by stimulus strength.

In contrast, the nARC (Fig.~\ref{fig3}A2) reveals a more strongly nonlinear and phase-specific modulation of oscillation amplitude. Pulses delivered at intermediate phases produce a pronounced suppression of amplitude (negative nARC), whereas stimulation at early and late phases tends to enhance oscillation amplitude (positive nARC). This pattern is robust across pulse intensities but becomes more pronounced for stronger inputs, particularly at 225 pA and 250 pA, indicating that amplitude modulation is both phase- and intensity-dependent. Notably, the structure of the nARC does not simply mirror the nPRC, suggesting that phase resetting and amplitude modulation are partially dissociable dynamical processes.

%%%%%%%%%%%%%%%%%%%%%%%%%% Fig3
%%%%%%%%%%%%%%%%%%%%%%%%%%

Fig.~\ref{fig3}B, depicting the change in pFF (defined as $\rm \Delta pFF = pFF_{Perturbed} - pFF_{Baseline}$), provides additional insight into how single pulses influence network synchrony. Here, negative values of $\Delta \mathrm{pFF}$ indicate that the pFF of the perturbed network is lower than baseline, corresponding to reduced synchrony, whereas positive values indicate an increase in pFF, reflecting enhanced synchrony. The $\Delta \mathrm{pFF}$ traces exhibit a pronounced phase dependence, with stimulation at intermediate phases ($\varphi = 0.4$-$0.6$) producing the largest decrease in pFF, indicating transient desynchronization of population activity. In contrast, stimulation at later phases tends to increase pFF, reflecting enhanced synchronization. These results emphasize that the network's response to external perturbations is highly state-dependent, revealing a rich dynamical landscape in which phase, amplitude, and synchrony are differentially modulated by transient inputs. More generally, they demonstrate that the timing of external stimuli can modulate not only the phase and amplitude of the ongoing network oscillations but also the degree of collective synchronization, providing a mechanism for phase-specific stabilization or destabilization of oscillatory activity in excitatory-inhibitory networks.

%%%%%%%%%%%%%%%%%%%%%%%%%%%%%%%%%%%%%%%%%%%%%%%%%%%%%

\subsection{Phase-dependent modulation of synchrony}

\begin{figure}[t!]
\centering
\includegraphics[scale = 0.55]{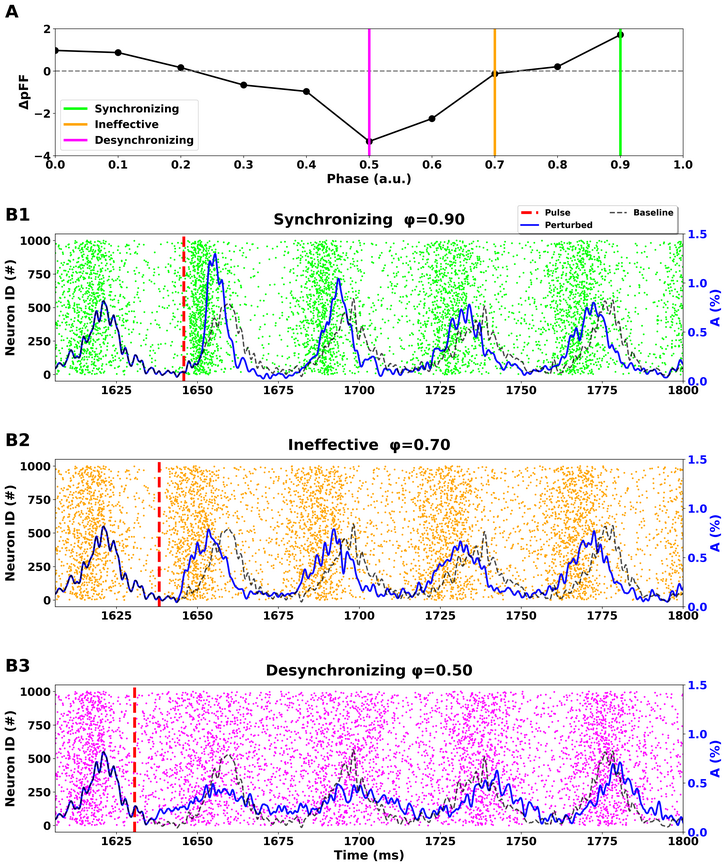}
\caption{{\bf Phase-dependent effects of single-pulse stimulation on network synchrony.} (\textbf{A}) Change in pFF ($\Delta \mathrm{pFF}$) as a function of stimulation phase for a 225 pA pulse. Three representative phases were selected: a synchronizing phase ($\varphi = 0.90$, green vertical line with $\Delta \mathrm{pFF}>0$), an ineffective phase ($\varphi = 0.70$, orange vertical line with $\Delta \mathrm{pFF} \approx 0$), and a desynchronizing phase ($\varphi = 0.50$, pink vertical line with $\Delta \mathrm{pFF}<0$). The gray dashed line indicates zero change. (\textbf{B1-B3}) Raster plots (colored spikes) and population activities (blue solid trace) of the perturbed network compared with the baseline, unperturbed network (gray dashed trace). The red dashed line marks pulse delivery. Depending on stimulation phase, the same pulse can enhance (B1), minimally affect (B2), or disrupt (B3) network synchrony.}
\label{fig4}
\end{figure}

To further illustrate the consequences of phase-dependent network response to stimulation, in Fig.~\ref{fig4} we examined three representative phases (marked by vertical lines with different colors) for a pulse intensity of 225 pA as an example. These phases were selected based on their distinct effects on network synchrony as quantified by $\Delta \mathrm{pFF}$: a synchronizing phase ($\varphi = 0.90$), an ineffective phase ($\varphi = 0.70$), and a desynchronizing phase ($\varphi = 0.50$). Fig.~\ref{fig4}A summarizes the phase dependence of $\Delta \mathrm{pFF}$ for this stimulation intensity (225 pA), revealing regions of enhanced synchrony ($\Delta \mathrm{pFF}>0$) at early and late phases of the oscillation cycle and a broad region of reduced synchrony ($\Delta \mathrm{pFF}<0$) centered around intermediate phases of oscillation (as demonstrated in Fig.~\ref{fig3}B).

%%%%%%%%%%%%%%%%%%%%%%%%%% Fig4
%%%%%%%%%%%%%%%%%%%%%%%%%%

\begin{itemize}

\item{\textit{Synchronizing stimulation:}} The raster plots and population activities in Fig.~\ref{fig4}B1-B3 demonstrate that the same stimulation pulse can produce qualitatively different network responses depending solely on its timing. When delivered at the synchronizing phase ($\phi=0.90$; Fig.~\ref{fig4}B1, red dashed line), the pulse induces a pronounced and coherent population (solid blue trace) event immediately following stimulation. Compared with the baseline activity (gray dashed trace), the perturbed network exhibits a sharper and larger population burst, indicating that a greater fraction of neurons are recruited into a temporally aligned response. This enhanced coordination persists for several oscillation cycles, leading to an increase in pFF relative to baseline and thus an increase in network synchrony ($\Delta \mathrm{pFF}>0$).

\item{\textit{Ineffective stimulation:}} In contrast, stimulation at the ineffective phase ($\phi=0.70$; Fig.~\ref{fig4}B2) produces only minor deviations (temporal shifts) from the ongoing network dynamics. The perturbed population activity (blue solid trace) closely follows the baseline trajectory (gray dashed trace), with only small transient differences in the timing and magnitude of population bursts. Consequently, the overall synchrony of the network remains largely unchanged, consistent with the near-zero value of $\Delta \mathrm{pFF}$ observed at this phase ($\Delta \mathrm{pFF} \approx 0$).

\item{\textit{Desynchronizing stimulation:}} A markedly different response emerges when the pulse is delivered at the desynchronizing phase ($\phi=0.50$; Fig.~\ref{fig4}B3). In this case, the stimulation disrupts the ongoing oscillatory organization of the network, resulting in weaker and more dispersed population bursts. The raster plot reveals a broader temporal distribution of spikes across neurons, indicating a loss of coordinated firing. Correspondingly, the population activity becomes less structured and exhibits reduced peak amplitudes compared with baseline (cf. Fig.~\ref{fig4}B3, blue and gray traces). This disruption increases the variability of collective activity, producing a negative $\Delta \mathrm{pFF}$ and therefore a reduction in synchrony ($\Delta \mathrm{pFF}<0$).

\end{itemize}

Taken together, these results demonstrate that the effect of stimulation cannot be predicted solely from its intensity. Instead, the phase of the underlying oscillation critically determines whether a pulse reinforces, leaves unchanged, or disrupts the ongoing collective dynamics in excitatory-inhibitory networks. Particularly, as demonstrated in Fig.~\ref{fig4}B3, the existence of desynchronizing windows within a single oscillatory cycle suggests that external stimuli can selectively steer network states through precise temporal targeting. Such phase-specific control may provide a mechanistic basis for designing stimulation protocols aimed at suppressing excessive neuronal synchrony in pathological oscillatory networks.

%%%%%%%%%%%%%%%%%%%%%%%%%%%%%%%%%%%%%%%%%%%%%%%%%%%%%

\subsection{Cumulative effects of desynchronizing stimulation}

Excessive neuronal synchronization is a hallmark of several neurological disorders, including Parkinson's disease and epilepsy~\cite{dominguez2005enhanced,hammond2007pathological,madadi2022inhibitory}. Consequently, considerable effort has been devoted to developing theory-based stimulation strategies capable of selectively disrupting excessive synchrony~\cite{popovych2014control}. The phase-dependent responses identified in Fig.~\ref{fig4} suggest that stimulation delivered at specific phases of the oscillation cycle can transiently desynchronize the network. By selectively delivering pulses only when they are effective (e.g., when pFF decreases), it is possible to desynchronize network activity.

A natural question is whether these transient effects accumulate over successive stimulation pulses to produce sustained changes in network dynamics. If the perturbation consistently disrupts the ongoing oscillation, repeated stimulation is expected to progressively weaken collective synchronization, leading to a gradual reduction in oscillation amplitude and the pFF. Such cumulative effects would indicate that the network responds predictably to repeated perturbations, providing a basis for phase-targeted desynchronization.

\begin{figure}[t!]
\centering
\includegraphics[scale = 0.28]{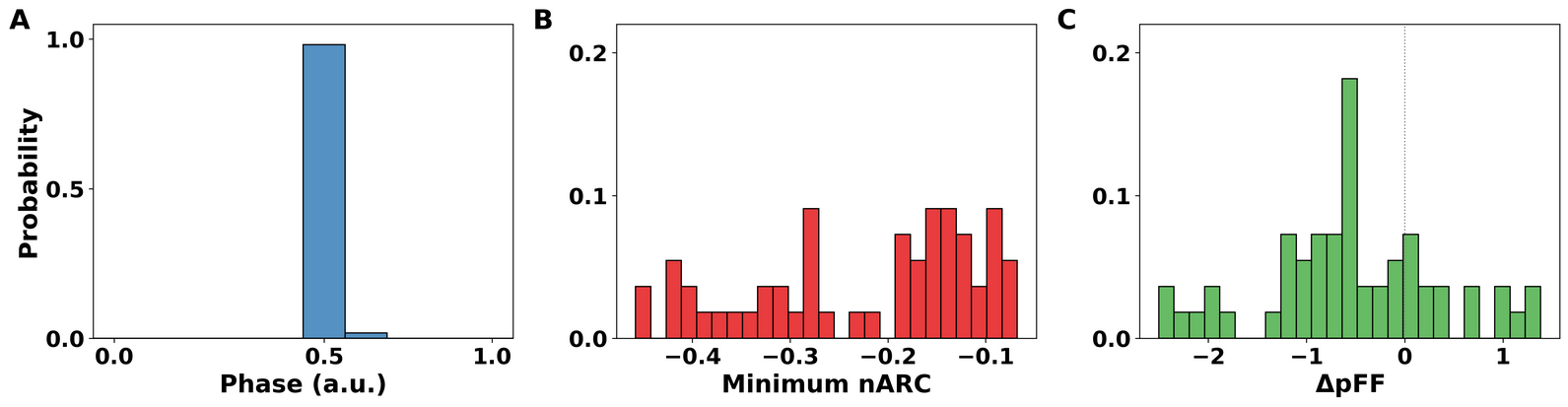}
\caption{{\bf Identification of an optimal phase for repeated desynchronizing stimulation.} (\textbf{A}) Distribution of the oscillation phases corresponding to the minimum nARC obtained for all pulse intensities (50-250 pA) and network realizations. The selected stimulation phase ($\varphi=0.5$) lies within the region where the largest negative amplitude responses consistently occur. (\textbf{B}) Distribution of the minimum nARC values corresponding to the selected phases, demonstrating that stimulation at these phases always reduces the oscillation amplitude. (\textbf{C}) Distribution of the change in pFF following a single stimulation pulse delivered at $\varphi=0.5$. The gray dashed line indicates zero change. Most realizations exhibit negative $\Delta\mathrm{pFF}$ values, indicating transient desynchronization of the network, although a small fraction show positive values because of variability across network realizations.}
\label{fig5}
\end{figure}

To address this question, we repeatedly applied stimulation pulses at the desynchronizing phase identified from the single-pulse analysis (Fig.~\ref{fig4}B3). Although phase resetting (nPRC) modifies the timing of the collective oscillation, it does not necessarily weaken the oscillation itself. In contrast, negative amplitude responses, as quantified by the nARC, directly reduce the strength of collective oscillatory activity and are therefore expected to more effectively disrupt excessive synchronization. To identify the optimal stimulation phase for desynchronization, we systematically examined the nARC over a wide range of pulse intensities (50-250 pA) and five network realizations. As shown in Fig.~\ref{fig5}, the minimum of the nARC consistently occurred within the phase interval $\varphi = 0.4$-$0.6$ (Fig.~\ref{fig5}A), and the corresponding amplitude responses were negative for all pulse stimulation intensities (Fig.~\ref{fig5}B). These results indicate that this phase interval constitutes a robust desynchronizing window, making it a suitable target for repeated stimulation aimed at progressively suppressing network oscillations.

\begin{figure}[t!]
\centering
\includegraphics[scale = 0.25]{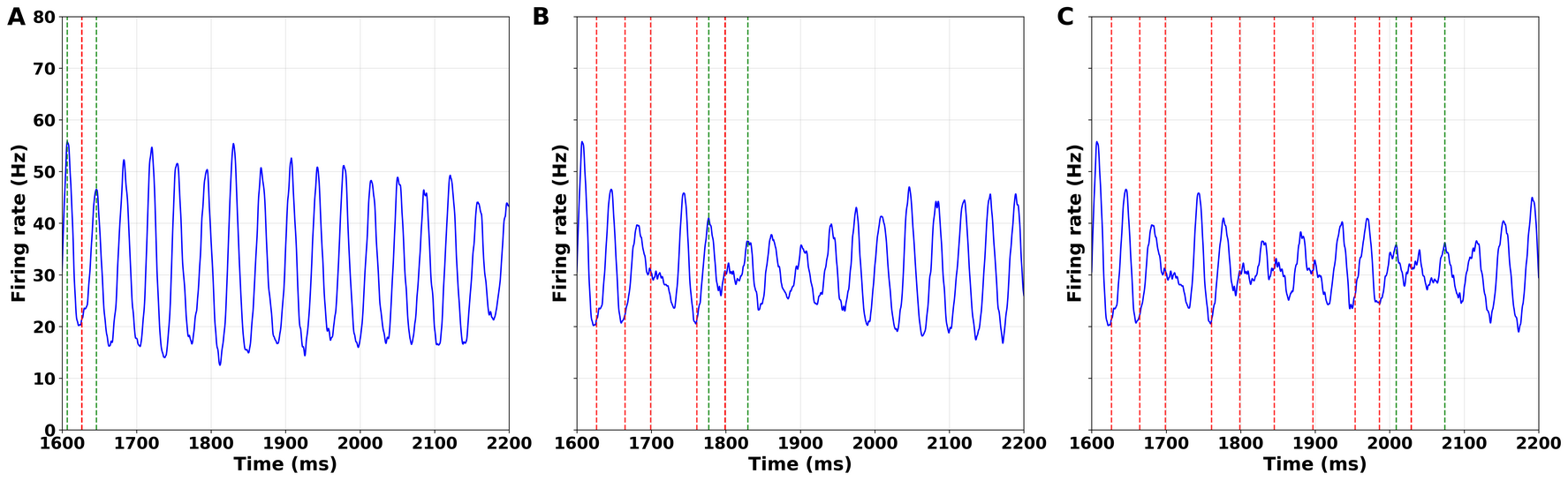}
\caption{{\bf Cumulative effects of repeated phase-targeted stimulation.} (\textbf{A-C}) Representative firing rate traces illustrating the progressive application of stimulation pulses (100 pA) at the desynchronizing phase ($\varphi=0.5$) in consecutive oscillation cycles; response following the first (A; 1-pulse), fifth (B; 5-pulse), and tenth (C; 10-pulse) stimulation pulses. As stimulation accumulates, subsequent pulses remain phase-locked to the oscillation while progressively reducing the amplitude and regularity of activity, consistent with cumulative desynchronization of the network. Red dashed vertical lines indicate the timing of stimulation pulses, while green dashed lines denote peaks of activity within the current analysis window.}
\label{fig6}
\end{figure}

Accordingly, we selected $\varphi = 0.5$ as the target phase for repeated desynchronizing stimulation. As an initial test, a single pulse (1-pulse) of varying intensity was delivered at this phase. For most stimulation amplitudes, the pulse successfully reduced network synchrony, as indicated by negative values of $\Delta \mathrm{pFF}$ (Fig.~\ref{fig5}C). However, a small number of cases exhibited positive $\Delta \mathrm{pFF}$, indicating a transient increase in synchrony, reflecting the stochastic variability arising from five different independent network realizations. We then investigated whether the desynchronizing effects could accumulate over time by repeatedly applying stimulation pulses at $\varphi = 0.5$ on 10 consecutive oscillation cycles. Representative examples of the network's firing rate response following the first (A; 1-pulse), fifth (B; 5-pulse), and tenth (C; 10-pulse) stimulation pulses are shown in Fig.~\ref{fig6} for one realization of the network.

%%%%%%%%%%%%%%%%%%%%%%%%%% Fig5
%%%%%%%%%%%%%%%%%%%%%%%%%%

%%%%%%%%%%%%%%%%%%%%%%%%%%% Fig6
%%%%%%%%%%%%%%%%%%%%%%%%%%

\begin{figure}[t!]
\centering
\includegraphics[scale = 0.35]{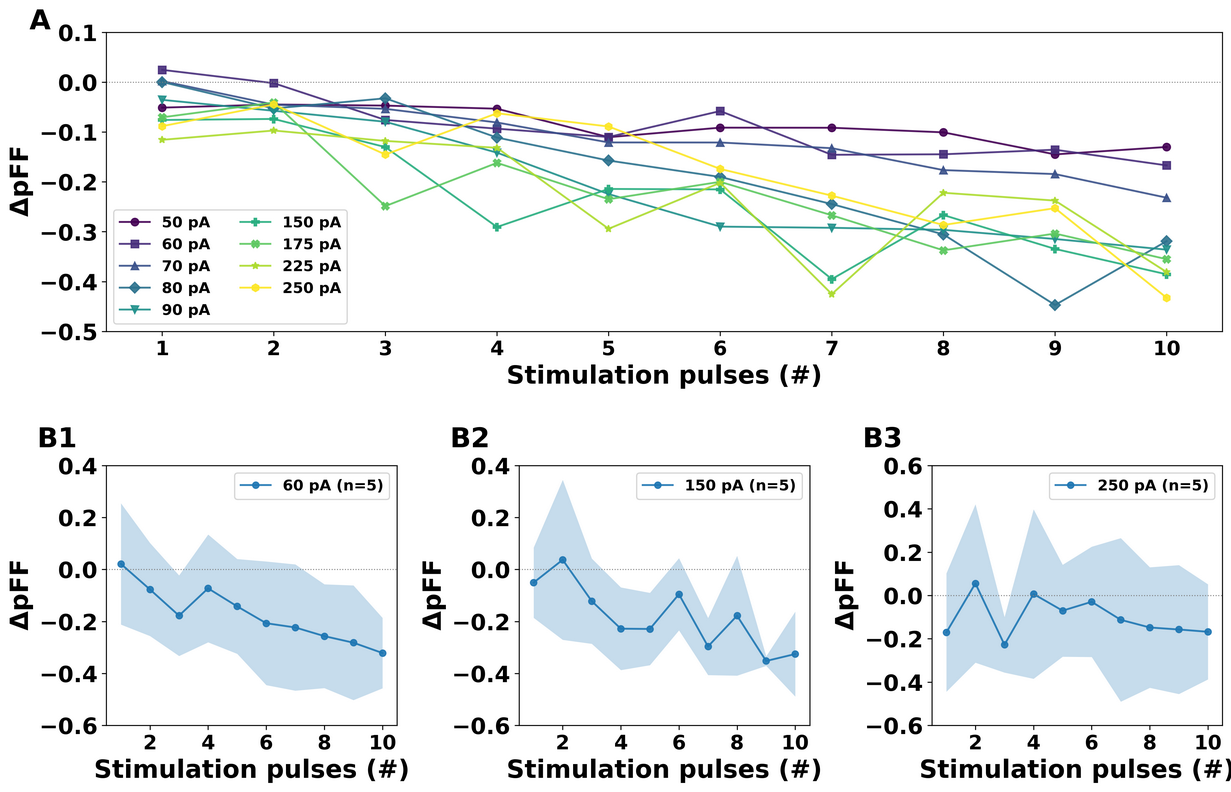}
\caption{{\bf Progressive desynchronization induced by repeated phase-targeted stimulation.} (\textbf{A}) Change in the pFF ($\Delta\mathrm{pFF}$) as a function of the number of stimulation pulses for different pulse intensities (50-250 pA, indicated by different colors and markers). Stimulation pulses were delivered at the optimal desynchronizing phase ($\varphi = 0.5$) in consecutive oscillation cycles. For all stimulation intensities, repeated stimulation progressively decreases $\Delta\mathrm{pFF}$, demonstrating the cumulative reduction of network synchrony. (\textbf{B1-B3}) Mean $\Delta\mathrm{pFF}$ for three representative stimulation intensities (60 pA, 150 pA, and 250 pA, respectively) averaged across five independent network realizations ($n = 5$). Shaded regions represent the standard deviation across network realizations. The gray dashed lines indicate zero change. Despite variability between realizations, repeated stimulation consistently shifts $\Delta\mathrm{pFF}$ toward negative values, indicating that phase-targeted stimulation robustly and progressively desynchronizes the network.}
\label{fig7}
\end{figure}

Following the application of ten stimulation pulses, as demonstrated in Fig.~\ref{fig7}A for a representative network realization, $\Delta \mathrm{pFF}$ progressively decreases with increasing numbers of stimulation pulses across a wide range of stimulation intensities (indicated by different colors and markers). This gradual reduction indicates that phase-targeted stimulation effectively weakens population synchrony and that its desynchronizing effects accumulate over successive oscillation cycles. Fig.~\ref{fig7}B1-B3 summarizes the mean response and variability of this effect for three representative pulse intensities (60 pA, 150 pA, and 250 pA) across five independent network realizations. Despite variability across individual realizations, repeated stimulation consistently reduced $\Delta \mathrm{pFF}$, demonstrating that appropriately timed perturbations can robustly disrupt ongoing synchronous activity. These findings suggest that repeated phase-specific stimulation progressively weakens collective synchronization, leading to a sustained reduction in population-level firing variability as quantified by the pFF.

%%%%%%%%%%%%%%%%%%%%%%%%%%% Fig7
%%%%%%%%%%%%%%%%%%%%%%%%%%

%%%%%%%%%%%%%%%%%%%%%%%%%%%%%%%%%%%%%%%%%%%%%%%%%%%%%

\subsection{Network response during repeated stimulation}

\begin{figure}[t!]
\centering
\includegraphics[scale = 0.3]{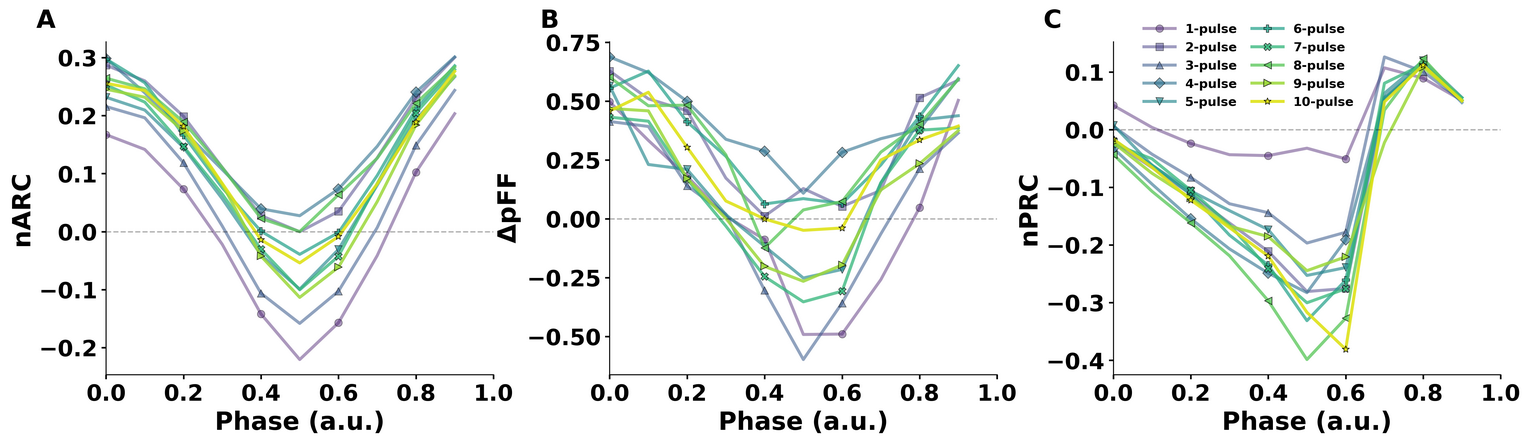}
\caption{{\bf Network responses during repeated phase-targeted stimulation.} Phase-dependent nARC (\textbf{A}), $\Delta\mathrm{pFF}$ (\textbf{B}), and nPRC (\textbf{C}) following one (1-pulse) to ten (10-pulse) consecutive stimulation pulses. Curves correspond to successive stimulation pulses and are shown for different stimulation intensities (averaged over 50-250 pA; for variations see Fig.~\ref{fig9}). The gray dashed lines indicate zero response or change. Although the characteristic phase dependence of the responses is preserved throughout the stimulation sequence, repeated stimulation progressively modifies their magnitude, reflecting the cumulative evolution of the network toward a less synchronized oscillatory state.}
\label{fig8}
\end{figure}

Repeated stimulation not only produces cumulative desynchronization but also progressively modifies the network's dynamical response to subsequent perturbations. Fig.~\ref{fig8} illustrates the evolution of the nARC, synchrony ($\Delta\mathrm{pFF}$), and nPRC following one to ten consecutive stimulation pulses (shown by different colors and markers) for different stimulation intensities (without showing variations). Despite quantitative differences across pulse amplitudes, all three response measures exhibit remarkably consistent phase-dependent profiles throughout the stimulation sequence, indicating that the underlying response characteristics of the network are largely preserved. However, the magnitude of these responses changes systematically with increasing numbers of stimulation pulses, demonstrating that repeated perturbations continuously reshape the dynamical state of the oscillatory network.

The most pronounced cumulative effect is observed for the nARC (Fig.~\ref{fig8}A). Although the characteristic U-shaped profile remains largely unchanged, the magnitude of the negative amplitude response gradually decreases as additional stimulation pulses are applied, particularly around the desynchronizing phase ($\varphi \approx 0.5$). This trend indicates that the oscillation becomes progressively weaker, leaving less oscillatory activity to be further suppressed by subsequent perturbations. A similar evolution is observed for $\Delta\mathrm{pFF}$ (Fig.~\ref{fig8}B). During the first few stimulation pulses, the network exhibits large reductions in synchrony near the optimal stimulation phase. As stimulation continues, however, the magnitude of $\Delta\mathrm{pFF}$ becomes progressively smaller, suggesting that the network approaches a new dynamical state in which synchrony has already been substantially reduced. Consequently, additional pulses produce diminishing incremental changes despite maintaining the network in a desynchronized regime.

The cumulative stimulation also modifies the network phase response (Fig.~\ref{fig8}C). While the overall shape of the nPRC remains similar across repeated stimulation, the negative phase-shift region around $\varphi \approx 0.5$ becomes progressively more pronounced before approaching a stable profile. This observation indicates that repeated stimulation not only suppresses oscillation amplitude but also alters the susceptibility of the network to phase resetting. Together, these results demonstrate that repeated phase-targeted stimulation drives the network toward a modified oscillatory state in which amplitude, synchrony, and phase sensitivity are jointly reorganized.

\begin{figure}[t!]
\centering
\includegraphics[scale = 0.37]{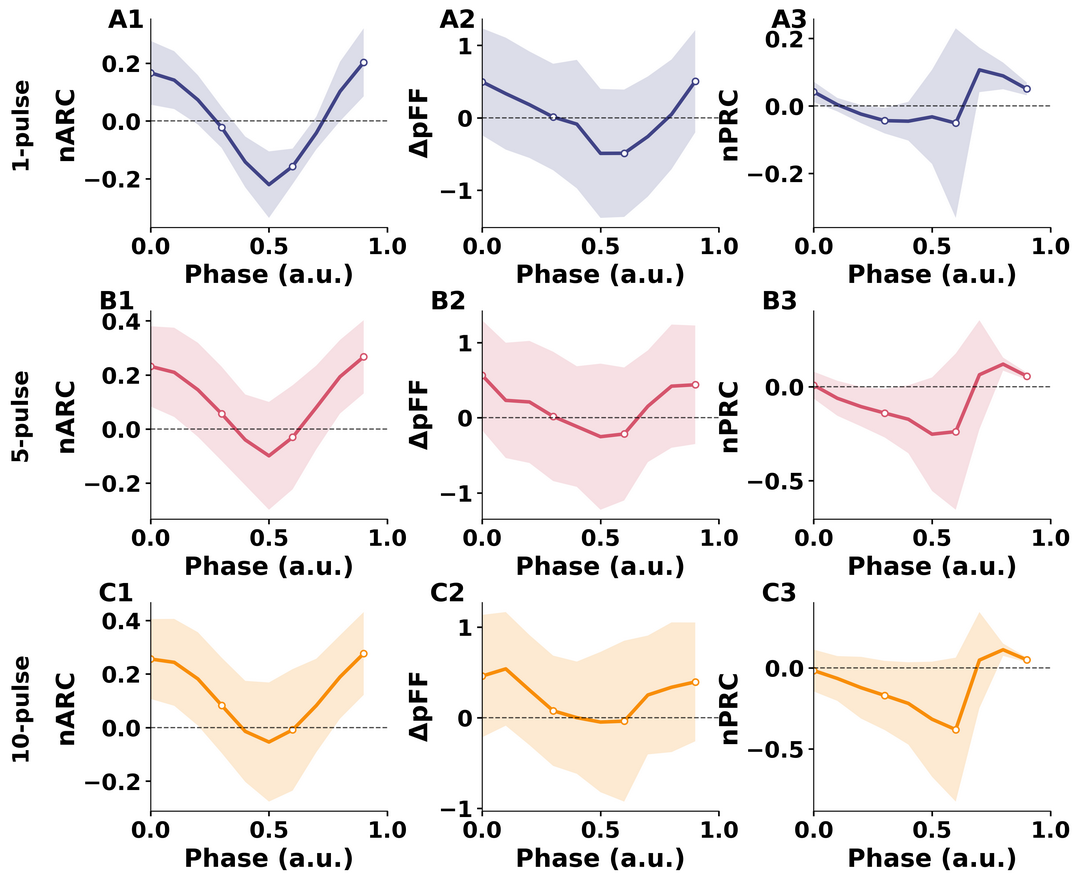}
\caption{{\bf Representative cumulative network responses following repeated stimulation.} Mean phase-dependent responses averaged across all stimulation intensities after one (\textbf{A1-A3}), five (\textbf{B1-B3}), and ten (\textbf{C1-C3}) consecutive stimulation pulses. Columns show the nARC, $\Delta\mathrm{pFF}$, and nPRC, respectively. Shaded regions indicate the standard deviation across stimulation intensities (50-250 pA). The gray dashed lines indicate zero response or change. The characteristic phase dependence is preserved throughout repeated stimulation, with the strongest desynchronizing effects consistently occurring near $\varphi \approx 0.5$, demonstrating that repeated stimulation weakens synchronization without substantially altering the optimal stimulation phase.}
\label{fig9}
\end{figure}

Fig.~\ref{fig9} presents the mean responses averaged across all pulse amplitudes for three representative stimulation pulses, i.e., 1-pulse (A1-A3), 5-pulse (B1-B3), and 10-pulse (C1-C3) stimulation, with shaded regions indicating the corresponding standard deviations. Averaging across stimulation strengths reveals that the characteristic phase dependence of the nARC, $\Delta\mathrm{pFF}$, and nPRC is highly conserved throughout the stimulation sequence. The nARC consistently exhibits a pronounced minimum near $\varphi \approx 0.5$, while $\Delta\mathrm{pFF}$ reaches its most negative values within the same phase interval, confirming that this phase remains the optimal target for desynchronizing stimulation even after repeated perturbations. Likewise, the nPRC preserves its biphasic structure, although the magnitude of the phase delay around the desynchronizing phase increases slightly with repeated stimulation.

Importantly, the relatively small changes in the mean response curves between one, five, and ten stimulation pulses, together with the substantial overlap of the standard deviations, indicate that repeated stimulation does not fundamentally alter the phase dependence of the network dynamics. Instead, it primarily changes the operating point of the oscillatory state, progressively weakening synchronization while preserving the phase window in which stimulation is most effective. These findings demonstrate that the cumulative desynchronizing effect arises from the gradual evolution of the network toward a less synchronized state rather than from a shift in the optimal stimulation phase, thereby supporting the use of a fixed phase-targeted stimulation protocol over prolonged stimulation periods.

%%%%%%%%%%%%%%%%%%%%%%%%%%% Fig8
%%%%%%%%%%%%%%%%%%%%%%%%%%

%%%%%%%%%%%%%%%%%%%%%%%%%%% Fig9
%%%%%%%%%%%%%%%%%%%%%%%%%%

%%%%%%%%%%%%%%%%%%%%%%%%%%%%%%%%%%%%%%%%%%%%%%%%%%%%%

\subsection{Robustness of cumulative desynchronization}

To assess the robustness of the proposed stimulation strategy, we examined whether the cumulative desynchronizing effects of stimulation persisted under variations in key model parameters. Specifically, we evaluated the network response over a broad range of pulse stimulation intensities (50-250 pA) and inhibitory synaptic decay time constants ($\tau^{\rm in}_{\rm syn}$ = 5-7 ms), the latter being a critical determinant of excitatory-inhibitory network dynamics. For each model parameter set, simulations were repeated across five independent network realizations, yielding 7,560 data points for the pulse stimulation intensity analysis and 500 data points for the inhibitory synaptic time constant analysis.

\begin{figure}[t!]
\centering
\includegraphics[scale = 0.35]{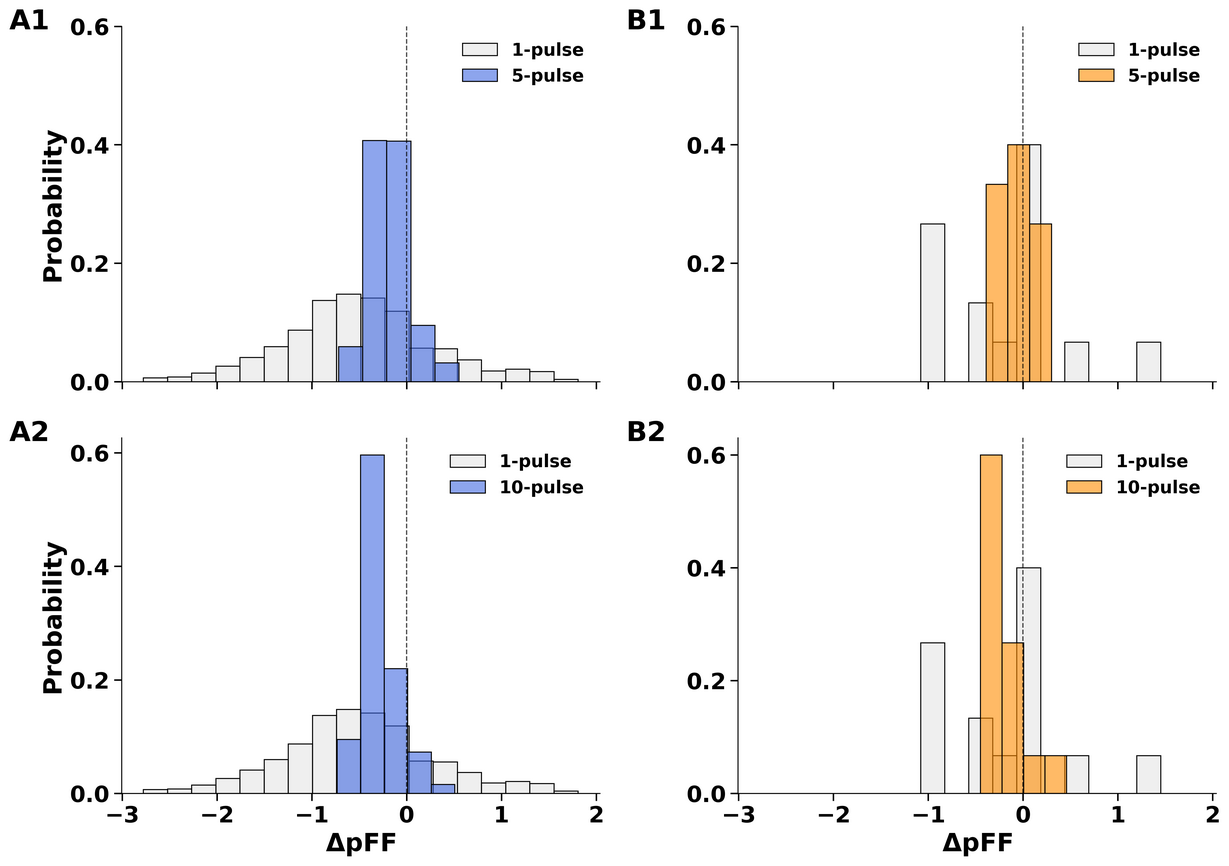}
\caption{{\bf Robustness of cumulative desynchronization under parameter variations.} Distribution of the change in pFF ($\Delta\mathrm{pFF}$) following single-pulse (1-pulse; gray background histograms) and repeated phase-targeted stimulation. (\textbf{A1, A2}) Comparison of single-pulse stimulation with five-pulse (5-pulse; blue, A1) and ten-pulse (10-pulse; blue, A2) stimulation across pulse intensities ranging from 50 to 250 pA. (\textbf{B1, B2}) Corresponding comparison for inhibitory synaptic decay time constants (5-7 ms; orange) across five independent network realizations. The vertical dashed lines indicate $\Delta\mathrm{pFF}=0$. Repeated stimulation progressively shifts the distributions toward negative $\Delta\mathrm{pFF}$ values while reducing their spread, demonstrating that cumulative desynchronization is robust to variations in stimulation intensity, inhibitory synaptic time constant, and network realization.}
\label{fig10}
\end{figure}

Fig.~\ref{fig10} compares the distributions of $\Delta\mathrm{pFF}$ following a single stimulation pulse (1-pulse; gray background histograms) with those obtained after five (5-pulse) and ten (10-pulse) consecutive stimulation pulses (colored histograms). For single-pulse stimulation, the distributions span a relatively broad range of values, including both negative and positive $\Delta\mathrm{pFF}$, indicating that the effect of an isolated perturbation depends on the specific network realization and parameter combination. Following five stimulation pulses (5-pulse; Fig.~\ref{fig10}A1 and B1), the distributions become noticeably narrower and shift toward negative values for both stimulation intensity (A1) and inhibitory synaptic time constant (B1). This shift indicates that repeated phase-targeted stimulation progressively suppresses population synchrony while reducing the variability of the excitatory-inhibitory network response.

After ten consecutive stimulation pulses (10-pulse; Fig.~\ref{fig10}A2 and B2), the desynchronizing effect becomes even more pronounced. Compared with the single-pulse case (1-pulse), the distributions are further concentrated around negative $\Delta\mathrm{pFF}$ values, whereas occurrences of positive $\Delta\mathrm{pFF}$ become increasingly rare. Importantly, this behavior is observed for both classes of parameter variation, demonstrating that the cumulative desynchronizing effect is largely insensitive to changes in stimulation intensity, inhibitory synaptic kinetics, and random network realization. Together, these results indicate that repeated stimulation not only enhances the average desynchronizing effect but also improves its reliability, making the proposed protocol robust to biologically relevant sources of variability.

%%%%%%%%%%%%%%%%%%%%%%%%%%% Fig10
%%%%%%%%%%%%%%%%%%%%%%%%%%

%%%%%%%%%%%%%%%%%%%%%%%%%%%%%%%%%%%%%%%%%%%%%%%%%%%%%
%%%%%%%%%%%%%%%%%%%%%%%%%%%%%%%%%%%%%%%%%%%%%%%%%%%%%

\section{Discussion}

The present study investigated how transient external perturbations reshape the collective dynamics of a balanced excitatory-inhibitory network exhibiting self-sustained oscillations. By jointly analyzing the nPRC, nARC, and $\Delta\mathrm{pFF}$, we characterized three complementary aspects of the network response: phase resetting, amplitude modulation, and synchronization. This multidimensional framework revealed that stimulation timing relative to the ongoing oscillation is the primary determinant of network behavior. Identical stimulation pulses produced qualitatively different outcomes, ranging from enhanced synchronization to pronounced desynchronization, solely as a consequence of the phase at which they were delivered. These findings demonstrate that collective network responses cannot be adequately described by a single response measure but instead emerge from the interaction of multiple dynamical dimensions in oscillatory networks.

A central finding of this work is that phase resetting alone provides an incomplete description of the response of oscillatory neuronal networks. Although the nPRC quantifies how perturbations advance or delay the collective rhythm, it does not necessarily predict whether stimulation weakens or strengthens the oscillation. In contrast, the nARC directly captures changes in oscillation amplitude and therefore provides a more direct indicator of whether external perturbations suppress or reinforce collective activity~\cite{bahadori2023efficient}. When interpreted together with changes in pFF, these measures distinguish perturbations that merely shift oscillation timing from those that genuinely alter the strength and synchronization of the network. Our results therefore support the use of integrated phase-amplitude-synchrony analyses when characterizing collective responses and designing stimulation strategies.

The phase dependence of the stimulation response further highlights the state-dependent nature of collective network dynamics and may have implications for phase-targeted neuromodulation. Within a single oscillation cycle, we identified distinct synchronizing, desynchronizing, and relatively ineffective stimulation windows, demonstrating that the same perturbation can produce opposite dynamical outcomes depending solely on its timing. These phase windows arise because network excitability varies continuously throughout the oscillation cycle as excitation and inhibition evolve. Similar phase-dependent modulation has been reported experimentally and computationally in studies of cortical and basal ganglia oscillations~\cite{cagnan2017stimulating,holt2019phase,duchet2020phase,west2022stimulating}, where stimulation delivered at appropriate phases selectively enhances or suppresses rhythmic activity. Our results provide a mechanistic explanation for these observations by showing that different phases correspond to distinct dynamical susceptibilities of the network. From this perspective, the oscillation cycle contains regions where perturbations reinforce collective activity and regions where they disrupt it. Exploiting these phase-dependent sensitivities may therefore enable more efficient stimulation protocols than approaches based solely on stimulus intensity or frequency.

An important advance of the present study is the demonstration that the effects of stimulation accumulate over repeated perturbations~\cite{hauptmann2009cumulative}. Rather than repeatedly reproducing the same transient response, phase-targeted stimulation progressively reshaped the response landscape itself. While the characteristic phase dependence of the nPRC, nARC, and pFF remained remarkably stable, their magnitudes evolved systematically as stimulation accumulated, indicating that the network was driven toward a modified operating point characterized by weaker oscillations and reduced synchronization. The preservation of the optimal desynchronizing phase throughout this process suggests that repeated stimulation modifies the network state without fundamentally altering the underlying mechanisms governing its phase sensitivity. This distinction is important because it indicates that a fixed phase-targeted stimulation protocol can remain effective over prolonged stimulation periods despite continuous changes in the excitatory-inhibitory network dynamics. 

Furthermore, the cumulative desynchronizing effects proved to be highly robust. Across broad variations in stimulation intensity, inhibitory synaptic decay time constant, and independent network realizations, repeated stimulation consistently shifted the network toward lower pFF values while reducing the variability of the response. Thus, repeated stimulation not only strengthened the average desynchronizing effect but also increased its reliability by making the network response progressively less sensitive to parameter variability. From the perspective of nonlinear dynamics, these findings suggest that repeated phase-targeted perturbations drive the system toward a stable region of state space in which desynchronization becomes increasingly reproducible despite intrinsic stochasticity and heterogeneity.

While the model captures key mechanisms underlying phase-dependent modulation of synchrony, several limitations should be acknowledged. The present work focused on a randomly connected excitatory-inhibitory network with homogeneous neurons and synapses, and examined responses to global stimulation delivered simultaneously to all neurons. Neural heterogeneity and spatially localized inputs in biological networks, however, can shape dynamics and computation in the brain~\cite{gast2024neural,wu2025neural,dahmen2026heterogeneity}. Furthermore, the model incorporates uniform synaptic weights across the network. This simplification may affect the precise location of synchronizing/desynchronizing stimulation windows and therefore limit the quantitative generality of our results. However, the qualitative principles identified here , i.e., the existence of phase-specific stimulation windows and the dissociation between phase, amplitude, and synchrony responses, are expected to be robust features of oscillatory excitatory-inhibitory networks~\cite{ma2017review}. As such, the model provides a useful framework for predicting when external stimuli are likely to stabilize or destabilize synchronized activity.

Moreover, the network was intentionally simplified to isolate the mechanisms underlying phase-dependent control. In particular, all synaptic interactions were assigned a fixed conduction delay. Although this assumption is common in large-scale spiking network models and facilitates interpretation of the results, biological neuronal networks exhibit substantial heterogeneity in axonal conduction times, synaptic transmission delays, and dendritic processing times~\cite{cakan2014heterogeneous}. Previous studies have shown that distributed delays can profoundly influence synchronization, oscillation frequency, multistability, and the emergence of coherent network states~\cite{ernst1995synchronization,crook1997role,brunel2000dynamics,roxin2005role,madadi2018delay}. Delay heterogeneity may therefore alter the shape of the nPRC and nARC, shift the locations of synchronizing and desynchronizing phases, or even create additional dynamical regimes that are absent in networks with uniform delays~\cite{yu2016heterogeneous}. 

In conclusion, our results demonstrate that phase-specific stimulation provides a powerful means of controlling synchronization in oscillatory networks. By jointly analyzing phase resetting, amplitude modulation, and changes in synchrony, we identify distinct dynamical regimes in which perturbations can synchronize, desynchronize, or minimally affect network activity. These findings highlight the multi-dimensional nature of network responses to perturbations and provide a basis for developing stimulation strategies that exploit the intrinsic dynamics of neuronal populations to reset pathological oscillations~\cite{krause2022brain}.

%%%%%%%%%%%%%%%%%%%%%%%%%%%%%%%%%%%%%%%%%%%%%%%%%%%%%
%%%%%%%%%%%%%%%%%%%%%%%%%%%%%%%%%%%%%%%%%%%%%%%%%%%%%

\section*{CRediT Author Statement}

\textbf{Ehsan Ahmadi:} Methodology, Formal analysis, Visualization, Investigation, Writing - review \& editing. \textbf{Mojtaba Madadi Asl:} Methodology, Formal analysis, Visualization, Writing - original draft, Writing - review \& editing, Project administration. \textbf{Alireza Valizadeh:} Conceptualization, Methodology, Formal analysis, Writing - original draft, Writing - review \& editing, Supervision.

%%%%%%%%%%%%%%%%%%%%%%%%%%%%%%%%%%%%%%%%%%%%%%%%%

\section*{Declaration of Competing Interests}

The authors declare that the research was conducted in the absence of any commercial or financial relationships that could be construed as a potential conflict of interest.

%The authors declare no competing interests.

%%%%%%%%%%%%%%%%%%%%%%%%%%%%%%%%%%%%%%%%%%%%%%%%%

\section*{Funding}

No funding was received for conducting this study.

%%%%%%%%%%%%%%%%%%%%%%%%%%%%%%%%%%%%%%%%%%%%%%%%%

\section*{Data Availability}

All data used to produce the figures were generated via numerical simulations. The simulation code is publicly accessible at \url{https://github.com/ehsanahmadi0013/Desynchronizing-phase-stimulation}.

%All data used to produce the figures was generated via numerical simulations of the freely available code.
%All data generated or analysed during this study are included in this published article.

%%%%%%%%%%%%%%%%%%%%%%%%%%%%%%%%%%%%%%%%%%%%%%%%%%%%%
%%%%%%%%%%%%%%%%%%%%%%%%%%%%%%%%%%%%%%%%%%%%%%%%%%%%%

{\footnotesize \bibliography{references}}
\bibliographystyle{vancouver}
\addcontentsline{toc}{section}{References}

\end{document}